\documentclass[]{aa} 
\usepackage{graphicx}
\usepackage{natbib}
\usepackage[english]{babel}
\usepackage{url}
\bibpunct{(}{)}{;}{a}{}{,}
\usepackage{txfonts}
\begin{document}
\selectlanguage{english}
   \title{Global stellar variability study in the field-of-view of the \textit{Kepler} satellite}

   \author{
    J. Debosscher \inst{1}
   \and
     J. Blomme \inst{1}
   \and
     C. Aerts\inst{1,2}
   \and
     J. De Ridder\inst{1}
}
   \institute{Instituut voor Sterrenkunde, KU Leuven, Celestijnenlaan 200B, 3001 Leuven, Belgium
           \and
          Department of Astrophysics, IMAPP, Radboud University Nijmegen, POBox 9010, 6500 GL Nijmegen, the Netherlands}

   \date{}

 
  \abstract
   {}
   {We present the results of an automated variability analysis of the \textit{Kepler} public data measured in the first quarter (Q1) of the mission. In total, about 150\,000 light curves have been analysed to detect stellar variability, and to identify new members of known variability classes. We also focus on the detection of variables present in eclipsing binary systems, given the important constraints on stellar fundamental parameters they can provide.}
   {The methodology we use here is based on the automated variability classification pipeline which was previously developed for and applied successfully to the CoRoT exofield database and to the limited subset of a few thousand \textit{Kepler} asteroseismology light curves. We use a Fourier decomposition of the light curves to describe their variability behaviour and use the resulting parameters to perform a supervised classification. Several improvements have been made, including a separate extractor method to detect the presence of eclipses when other variability is present in the light curves. We also included two new variability classes compared to previous work: variables showing signs of rotational modulation and of activity.}
   {Statistics are given on the number of variables and the number of good candidates per class. A comparison is made with results obtained for the CoRoT exoplanet data. We present some special discoveries, including variable stars in eclipsing binary systems. Many new candidate non-radial pulsators are found, mainly $\delta$ Sct and $\gamma$ Dor stars. We have studied those samples in more detail by using 2MASS colours. The full classification results are made available as an online catalogue.}
   {}

   \titlerunning{Automated supervised classification of variable stars}
   \maketitle
%

\section{Introduction}

The NASA \textit{Kepler} mission has been operational for more than 1.5 years now (see \cite{Borucki-Kepler}, for a description of the mission and some first results). Its major science goal is the detection of exoplanets, and in particular Earth-like planets. Similar to the CoRoT mission \citep{COROTBOOK,Auvergne-COROT}, the transit method is used to detect signatures of exoplanets. This method requires the precise and continuous monitoring of large numbers of stars. As a consequence, a gold mine of variable star light curves at $\mu$mag precision is being produced. The nominal maximum time-span of the data will be 4 years, while this is only some 150 days for CoRoT. \textit{Kepler} will thus allow us to explore much longer time-scales than CoRoT, while CoRoT's denser time sampling (512s versus 29.4m) is more suited to probe the short-period variability domain.

We performed a global variability analysis of the public \textit{Kepler} Q1 data, which was released on 15 June 2010, using automated supervised classification and extractor methods. Statistics on the number of variables and estimates of the class populations are presented. Special attention is paid to the detection of eclipsing binaries, and in particular, pulsating stars in eclipsing binary systems. The latter can provide us with model independent constraints on astrophysical parameters such as mass and radius, which constitute essential input for asteroseismological studies.
Since we find a large number of new $\gamma$ Dor and $\delta$ Sct candidates, we investigate the observational properties of these samples in more details and compare them with those of known $\gamma$ Dor and $\delta$ Sct stars in order to see if the improved precision leads to an extension of the observational instability strips. We also evaluate the samples of objects assigned to the new rotational modulation and stellar activity classes now taken into account by our classifiers, after CoRoT provided us with appropriate light curves to define these two classes.

To conclude, we briefly present a comparison of some Kepler
light curves with ground-based TrES data of the same targets that
we analysed using similar methods.

The classification results are made available to the astronomical community in electronic form, since they are very useful for target selection and to study different statistical aspects of the \textit{Kepler} data.

\section{Data description}

The data analysed in this work include all $\sim$150\,000 public \textit{Kepler} light curves, measured in the first quarter of the mission. The total time span of the light curves is about 33 days, with a sampling interval of 29.4 minutes (long-cadence data). Only a small fraction of the light curves has been measured in short-cadence mode, where the sampling interval equals $\sim$1 minute. \textit{Kepler} is observing in white light, with a bandpass of 430-890nm FWHM. The observed stars have magnitudes ranging from 9 to 16.
We used the corrected fluxes for our analysis, since they suffer less from instrumental systematics, and most outliers have been removed from the data already. To complement the light curve information and to evaluate our results, we also use the 2MASS colour indices present in the KIC catalogue (\textit{Kepler} input catalogue).

\section{Methodology}

The methodology is similar to the one applied in \cite{Blomme-Kepler}, and described in more detail in \cite{COROT-JD}. Basically, we describe the main characteristics of each light curve by performing a Fourier-decomposition, including a maximum of $3$ independent frequencies, each with a number of overtones. The Fourier parameters are then fed to the supervised classifier, where they are compared to the parameters of template light curves (training set) belonging to several known stellar variability classes. Class assignment is done in a probabilistic way, since light curves can share characteristics of several variability classes at the same time. We keep improving the capabilities of the classifiers, and have now extended our training set to be able to recognize light curves showing the signs of rotational modulation and activity. We used the clustering results obtained with the CoRoT data, as presented in \cite{Sarro-clustering} to define these two new classes. Their template data consist of CoRoT exoplanet field light curves for now, but they will be extended in the future, since \textit{Kepler} will provide many new examples. The definition of these new classes is still somewhat experimental, but as will be shown further on, good results are obtained with both classes.

We have also extended the methods to improve the detection of (single-)eclipses in light curves, regardless of the presence of other variability. It concerns an automated extractor method which complements the results of our supervised classification. The method is described in more detail in the following section.

\subsection{Eclipsing binary detection}
\begin{figure*}
 \centering
   \includegraphics[width=16cm,angle=270,scale=0.8]{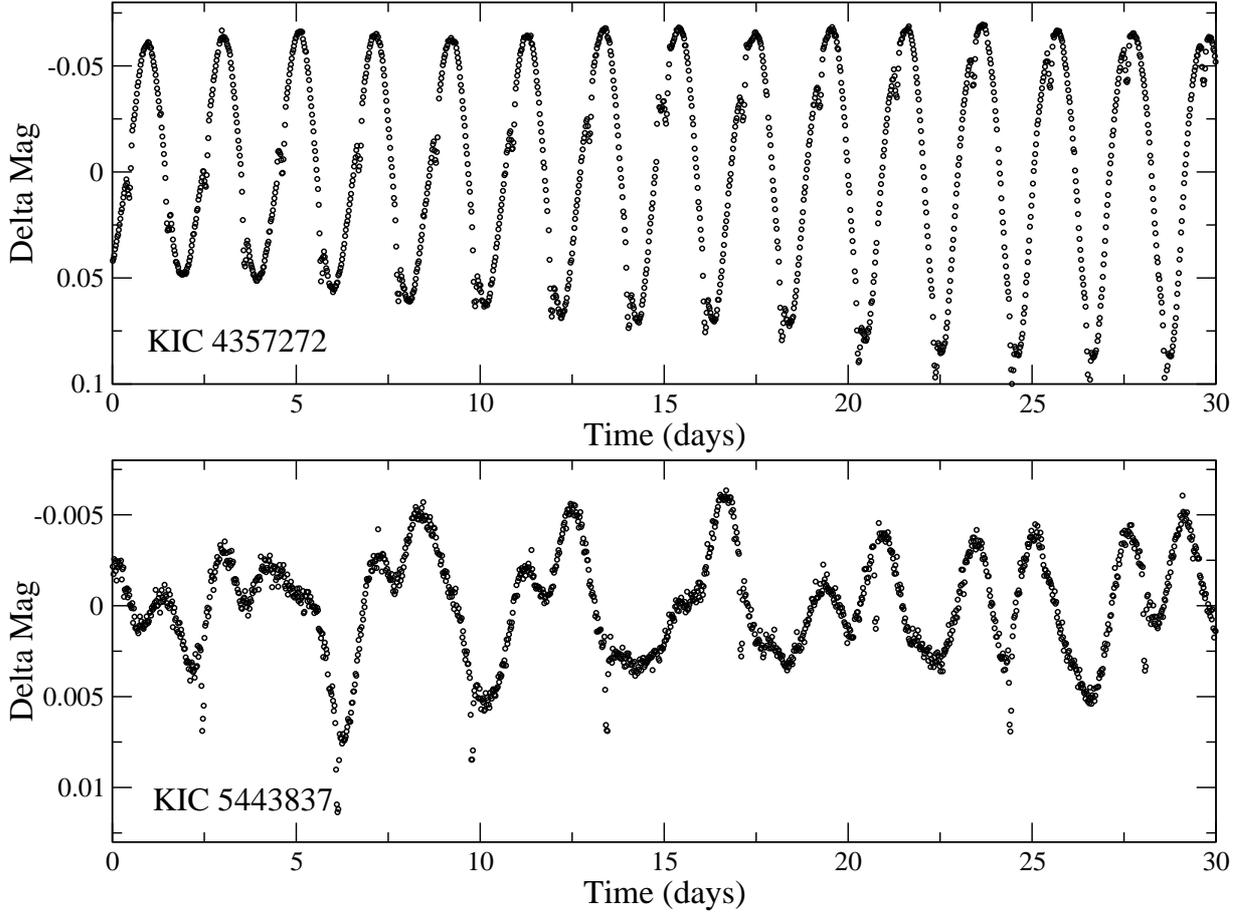}
\caption{Two examples of light curves showing eclipses, and detected with our dedicated extractor method. The presence of additional variability in these light curves (possibly due to spots) caused them to be missed as binaries by our regular classifiers.}
\label{ecl-detect}
\end{figure*}

\begin{figure*}
 \centering
   \includegraphics[width=16cm,angle=270,scale=0.8]{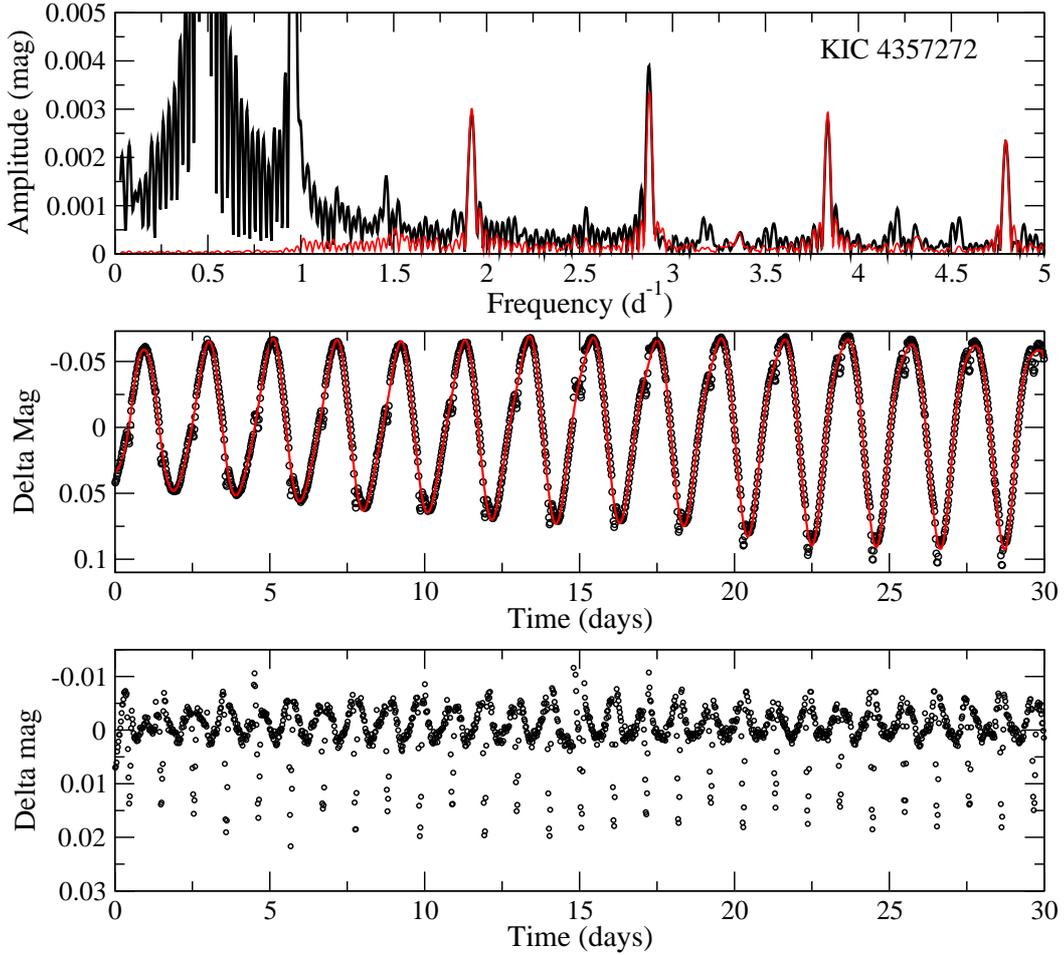}
\caption{Filtering process illustrated for KIC 4357272. The top plot shows the amplitude spectrum before (in black) and after the high-pass filtering (in red). The signal below the cutoff-frequency of 1.5 $d^{-1}$ has been completely removed, while the signal at higher frequencies is retained, and now has a lower noise level. The middle plot shows the original light curve (black circles) with its low-frequency part superimposed (red curve, signal below 1.5 $d^{-1}$). The high-pass filtered version is then obtained by subtracting the red curve from the original light curve, as shown in the bottom plot. The eclipses are now clearly visible and easily detectable with automated methods.}
\label{fil-example}
\end{figure*}

Our classifiers are able to identify eclipsing binaries in a reliable way, provided that several orbital periods are sampled by the light curve, or that a sufficient amount of measurements during eclipse is present. Otherwise, their signatures in the Fourier spectrum are very weak and difficult to identify with an automated method. Those cases are likely to be missed by the classifier. The presence of additional variability in the light curve, either instrumental or intrinsic to the object, hampers the detection of eclipses even more. Therefore, we have developed an extractor method for those cases, which effectively complements the other classifiers. This method also allows to detect eclipses when the orbital period of the binary is similar to or even equal to the period of the additional variability in the light curve. Basically, eclipses are detected as downward outliers in a high-pass filtered version of the light curve. The high-pass filtering removes the low-frequency content of the light curves, including instrumental trends and long-timescale variability. The resulting filtered version of the light curve only retains the high frequency content, including part of the highly non-sinusoidal eclipse signal (the higher harmonics of the orbital frequency). As an additional advantage, several combination frequency peaks are removed as well (e.g. combinations of low frequencies which are filtered out, and higher frequencies). This effectively makes the high frequency region in the amplitude spectrum less contaminated. The filter works by convolving the original light curve $y(t_i,i=1,N)$ with a sinc-function $k(t_i)$, resulting in a new light curve $Y(t_i)$ :

\begin{equation}
Y(t_i)=(y*k)(t_i)=\sum_{j=1}^N {y(t_j)*k(t_i-t_j)},
\end{equation}
where $k(t)$ is defined as:
\begin{equation}
k(t)=\frac{\sin(2 \pi f t)}{2 \pi f t},
\end{equation}
with $f$ the cutoff frequency, to be defined by the user. In our application to the Kepler data, we used a cutoff frequency of 1.5 $d^{-1}$. All frequencies above this value will be removed from the light curve.
This technique is well known in electronic filtering systems. It is based on the mathematical result that convolution with a sinc-function in the time domain corresponds to multiplication with a rectangular bandpass function in the Fourier domain. The resulting light curve $Y(t_i)$ is a low-pass filtered version of the original light curve $y(t_i)$. The desired high-pass filtered version $y_{hf}(t_i)$ is then obtained as:
\begin{equation}
y_{hf}(t_i)=y(t_i)-Y(t_i).
\end{equation}
We now scan $y_{hf}(t_i)$ for groups of downward outliers using box-plot statistics. This method has the advantage of being less sensitive to the underlying statistical distribution of the data. In the application to the \textit{Kepler} data, we flagged the light curve if more than 10 outliers were detected this way. This flag was then combined with the usual classification labels. Figure \ref{ecl-detect} shows two examples of eclipsing binaries detected using this method, while Fig. \ref{fil-example} illustrates the filtering process for one of the light curves. The filter will remove any kind of variability with frequencies below the cutoff value, but the eclipse detection only works well, if the additional variability (not related to the eclipses), is confined to a frequency region below the cutoff frequency of the filter, and if the eclipse signal has sufficient power (in the form of higher harmonics) above the cutoff frequency.

The value of the cutoff frequency we chose proved to be a good compromise between removing sufficient low-frequency content of the light curves (hampering the eclipse detection) and not removing too much high frequency content, since the latter contains part of the eclipse signal. Given that we are especially interested in detecting pulsators in eclipsing binary systems (see Sect. 4.3), in particular of $\gamma$ Dor and SPB type, this cutoff value will remove most of the pulsation signal for those targets, making the detection of eclipses possible in these cases. At the expense of computation time, different cutoff values can be tried and the eclipse detection can be done on different filtered versions of the light curves. For example, a higher cutoff frequency can be chosen to be able to detect eclipses in the presence of higher frequency pulsations (otherwise, the filter will not remove any `disturbing' variability). However, higher cutoff values also remove more of the eclipse signal, and not enough power might remain to detect them. To avoid this, and to limit the computation time, we performed an additional outlier detection step at the end of the light curve analysis procedure. The automated procedure removes a maximum of 3 different frequencies from the light curve (with each a maximum of 4 harmonics), in 3 consecutive prewhitening steps. This way, we filter out only the dominant signal, irrespective of its frequency. The residuals are then again checked for downward outliers, indicative of eclipses.
It is clear that a combination of techniques is needed to detect all kinds of eclipse signals, even more so because additional variability on several timescales can be present as well. Our regular classifier reliably detects `pure' eclipsing binary light curves, irrespective of their orbital period, and the extractor methods can detect eclipses in the presence of additional variability, or when the eclipse signal is too faint to cause clear signatures in the Fourier spectrum. The fact that the extractor scans the (filtered) light curve for outliers implies that it is well suited to detect detached system, with highly non-sinusoidal light curves. Close binaries, showing sinusoidal-like light curves, are better detected with the regular classifier.

\section{Classification results}
\subsection{Number of variables}
\begin{figure}
 \centering
   \includegraphics[width=16cm,angle=270,scale=0.4]{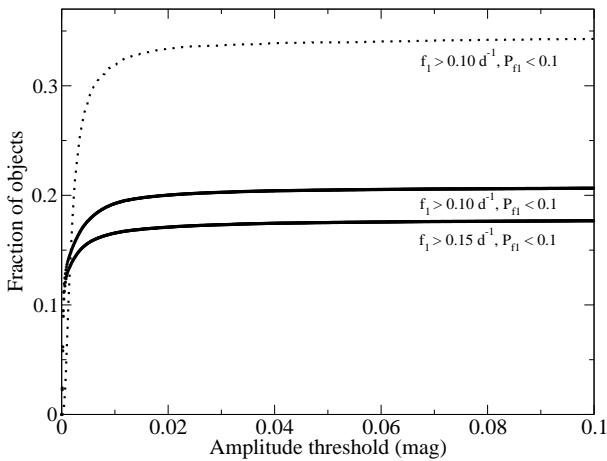}
\caption{Fraction of objects with $f_1\geq0.1,0.15$ $d^{-1}$, $P_{f_1}\leq0.1$, and having an amplitude below a certain threshold value, as a function of the threshold value (in magnitude). The dotted curve represents the results obtained for CoRoT \citep{COROT-JD}.}
\label{vars-amp}
\end{figure}

Following the application of our automated methods, we estimated the number of periodic variables in the dataset and constructed samples of good candidate members for the major stellar variability classes we included in our classifier. Variability estimates are listed in Table \ref{varfrac}, they can be compared to Table 4 in \cite {COROT-JD}, where we made the same estimates for the CoRoT exofield database. A detailed description of the variability selection criteria can be found there. In short, we take a light curve to be variable if at least one of the 3 highest peaks in the amplitude spectrum is significant (significance parameter $P_{f_i}<P_{max}$ ), and has a frequency value above a certain threshold ($f_i>f_{min}$). We list the resulting percentages for a few combinations of $f_{min}$ and $P_{max}$. If we compare these with a short CoRoT observing run having approximately the same time span as the \textit{Kepler} data, we find a significantly smaller fraction of variables. \textit{Kepler}'s noise levels per measurement are significantly lower, as shown in \cite{Blomme-Kepler}, but the time sampling is less dense: 29.4 min versus 8.5 min or even 32 sec for a significant fraction of the CoRoT data. Probably, the estimates for CoRoT, though conservative, were still influenced by instrumental effects, amongst other things caused by the passage through the South Atlantic Anomaly \citep{Auvergne-COROT}. This passage causes impacts of charged particles on the CCDs, influencing the pixel responses in several ways. Measured flux levels can temporary increase or decrease, and this translates to discontinuities in the light curves. We refer to e.g. \cite{COROT-jumps} for a more detailed description of these instrumental effects. Often, several discontinuities are present in a single CoRoT light curves, causing peaks in various regions of the amplitude spectrum, but always with significant power at frequencies below 0.15 $d^{-1}$. Figure \ref{vars-amp} plots the fraction of objects having significant variability (P-value of the dominant frequency $f_1$ below 0.1 ), and with a corresponding amplitude below a certain threshold, as a function of this threshold value. It is clear that the majority of variables have very low amplitudes, only reliably detectable using space-based instruments. This figure can be compared with Fig. 6 in \cite{COROT-JD}, where similar result were obtained (they are included in Fig. \ref{vars-amp}).
\begin{table}
\center
\caption{Fraction of light curves, fulfilling the criteria $f_i>f_{min}$ and $P_{f_i}<P_{max}$ for at least one of the 3 $f_i$'s, for four combinations of the thresholds $f_{min}$ (frequency threshold) and $P_{max}$ (significance threshold). For comparison, we also list the numbers for a CoRoT observing run of similar duration.}
 \begin{tabular}{l|c|c}
  \hline
  $f_{min}$, $P_{max}$& $\%$ of \textit{Kepler} objects&$\%$ of CoRoT objects\\
  \hline
 0.1 $d^{-1}$, 0.1& 28&35\\
 0.1 $d^{-1}$, 0.2& 32&40\\
 0.2 $d^{-1}$, 0.1& 20&29\\
 0.2 $d^{-1}$, 0.2& 24&34\\
\hline
\end{tabular}
\label{varfrac}
\end{table}
\subsection{Class statistics}
Table \ref{classes} summarizes the class statistics, including the remaining numbers of objects using different thresholds for the contamination level (taken from the KIC catalogue) of the light curves. We determined the number of good candidates for each class by first selecting the clearest variables assigned to each class using the criteria discribed in the previous section ($P_{f_1}\leq0.1$). Next we imposed limits on the Mahalanobis distance to the training class center for the remaining sample (similar to sigma clipping): We retained only those candidates having a Mahalanobis distance below 1.5. In short, this distance measure is a multi-dimensional generalization of the one-dimensional statistical or standard distance (e.g. distance to the mean value of a Gaussian in terms of sigma). This distance can effectively be used to retain only the objects that are not too far from the class centre in a statistical sense. More details on this distance measure can be found in \cite{COROT-JD}. Note that our classifiers take more variability classes into account than those listed in Table \ref{classes}. The full list of classes and their abbreviations, used by our current classifiers, can be found in Appendix A, while a description of the properties of the pulsators is available in Chapter 2 of \cite{Aerts-book}. We only list results here for those classes expected to be populated in the \textit{Kepler} field, and whose typical variability behaviour is detectable with the current time span of the light curves. Similar to CoRoT, the number of classical pulsators is small, owing to the \textit{Kepler} target selection procedure, which favoured mainly G-type stars on or near the main-sequence. 

No good Cepheid candidates have been identified, but some candidates might show up when longer datasets become avaliable. Of the few RR Lyr light curves we identified, the majority turned out to be heavily contaminated. In fact, they all showed the variability of RR Lyrae itself, which falls in the \textit{Kepler} field and whose brightness causes bleeding on the CCDs (Kolenberg, private communication). This illustrates that users of the \textit{Kepler} data must carefully check if their targets are contaminated or not. The presence of a neighbouring variable can introduce variability in the light curve of nearby targets, because part of the flux of the neighbouring variable might be included in the pixel mask. Note that more RR Lyr stars are present in the \textit{Kepler} observing field, but their light curves are not included in this public data release. They belong to the asteroseismology dataset, part of which was analysed by us in \cite{Blomme-Kepler}. Some first \textit{Kepler} results on those RR Lyr stars are described in \cite{Kolenberg-Kepler}.

\begin{figure*}
 \centering
   \includegraphics[width=16cm,angle=270,scale=0.9]{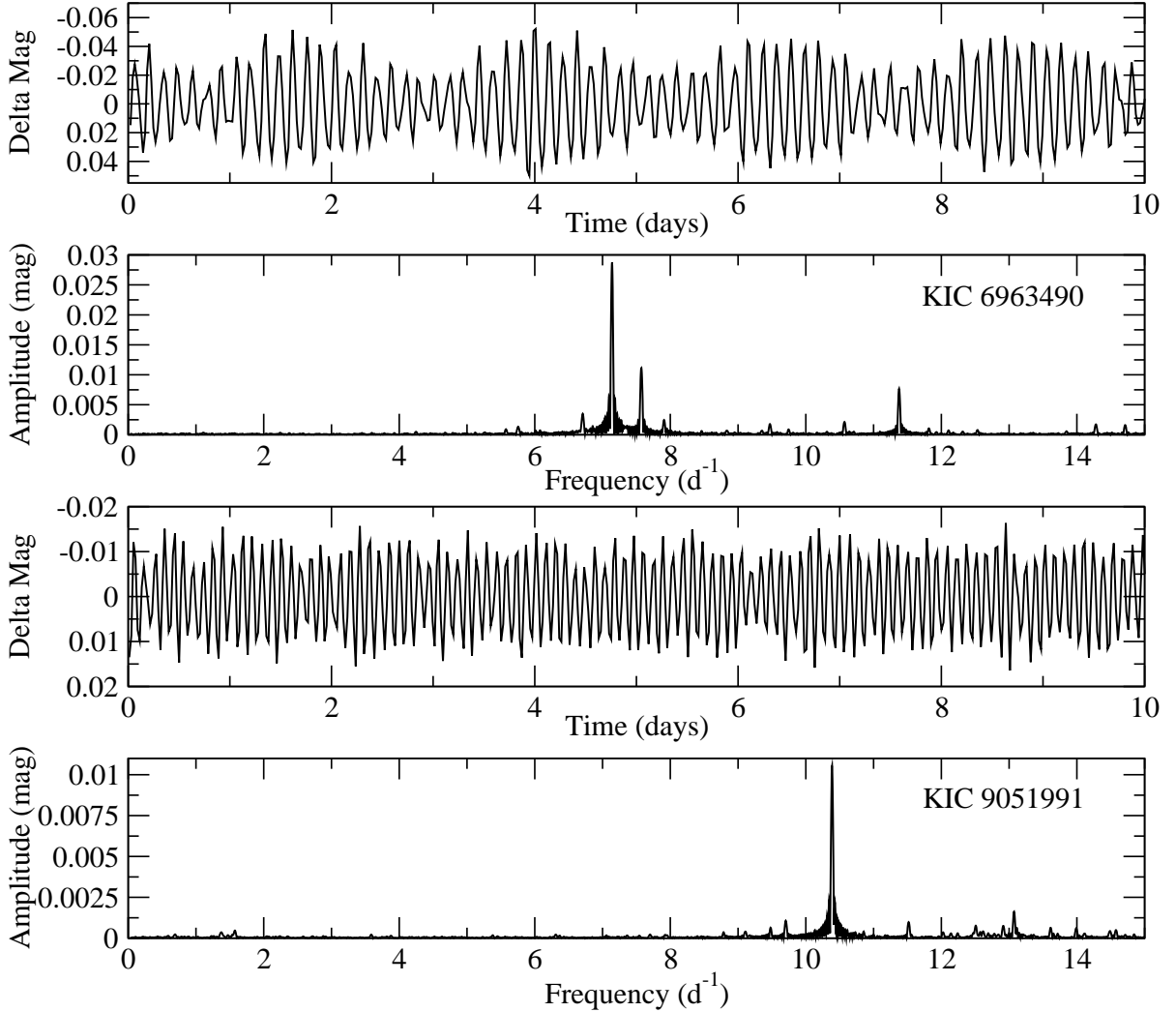}
\caption{Light curves and amplitude spectra of two objects in the binary list presented by \cite{Prsa-Kepler}, and classified by us as $\delta$ Sct stars.}
\label{dscut-examples}
\end{figure*}

\begin{table*}
\center
\caption{Major stellar variability classes and the number of good candidates we find for each in the public Q1 \textit{Kepler} data. Note that the binary category includes both eclipsing and ellipsoidal binaries.}
 \begin{tabular}{l|l|l|l}
  \hline
  Stellar variability classes& Candidates & Contamination $\leq$ 0.1&Contamination $\leq$ 0.01\\
  \hline
 RR-Lyrae stars, subtype ab&18&9&1\\
 RR-Lyrae stars, subtype c&4&2&1\\
 Delta-Sct/Beta-Cep&403&299&120\\
 Gamma-Doradus/SPB &441&304&92\\
 Binaries&2116&1105&262\\
 Stellar Activity and Rotational modulation& $>$3200&$>$1613&$>$370\\
\hline
\end{tabular}

\label{classes}
\end{table*}
Recently, a list of binaries in the \textit{Kepler} Q1 data was made available by \cite{Prsa-Kepler}. Since they focused on binaries only, and used dedicated methods to detect them, we compared our sample of binaries with their results. Their list contains 1879 objects, of which 1767 are present in the public Q1 dataset. Here, we only compared the results for the public light curves. We identified 1156 out of the 1767 objects as eclipsing binary or ellipsoidal variable using our global supervised classification method. The additional application of our dedicated extractor method increased the count to 1550 (88 $\%$), which is quite a good agreement given the very different nature of the methodology and the large diversity of variability classes we consider. We have manually checked the objects we did not recognize as binaries with either method (in total 217): about half of those are clearly eclipsing binaries and slipped our eclipse detection criteria, it concerns light curves with either very shallow eclipses, or some very uncommon light curves, not recognized by the current version of our classifier. Some 30 light curves have been confused with pulsators by our classifier: it concerns short period binaries with nearly sinusoidal light curves. They are confused with monoperiodic RR Lyr pulsators of subtype RRc or $\beta$ Cep pulsators. The true nature of the remaining half of the 217 light curves in the list is less clear without any additional information. About 10 of those show amplitude changes, indicative of rotational modulation, but the majority of the light curves resemble those of pulsating variables and are classified as such by our regular classifier. It is well known that light curves of close binaries can indeed be confused with those of RRc, high amplitude $\delta$ Sct and $\beta$ Cep pulsators. About 30 light curves are almost sinusoidal and the majority is classified by our methods as $\delta$ Sct or $\beta$ Cep. More investigation is needed to have certainty about those cases, but if some of them turn out to be pulsators, they are probably of RRc or $\delta$ Sct type (given that the massive  $\beta$ Cep stars are rare, see also Sect. 4.4).
Remarkably, about 40 of the 217 light curves clearly show multiperiodic pulsations and are classified by us as $\delta$ Sct stars with high probability. Visual inspection of those light curves and their amplitude spectra showed that it concerns clear $\delta$ Sct candidates indeed, as can be seen for two cases in Fig. \ref{dscut-examples}. We have also checked the orbital periods they list for those cases, and these turn out to be twice the value of the main pulsation period we detected in the amplitude spectrum.
\subsection{Variables in eclipsing binary systems}
\begin{figure*}
 \centering
   \includegraphics[width=16cm,angle=270,scale=0.7]{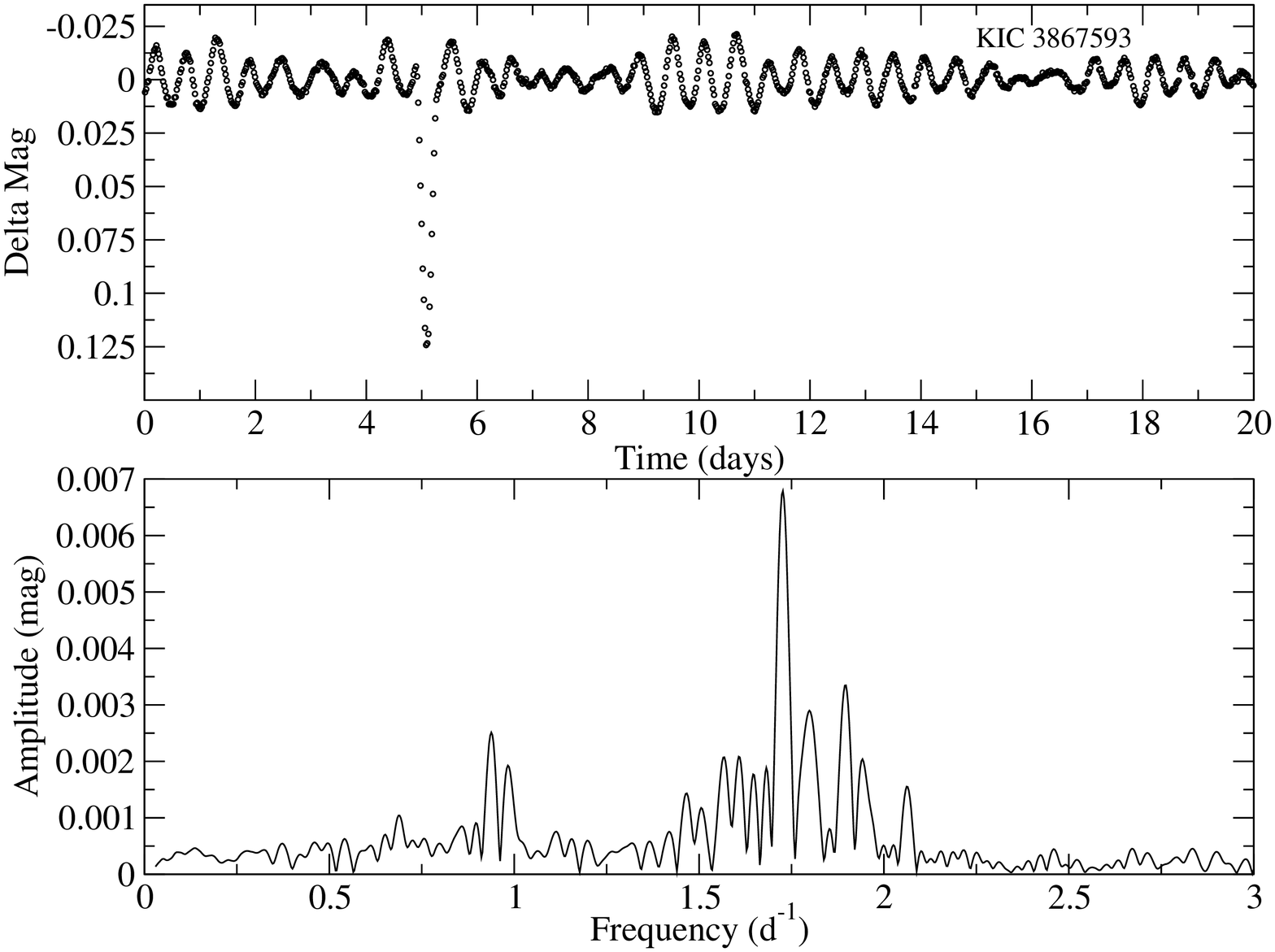}
\caption{An eclipsing binary containing a $\gamma$ Dor or SPB-type pulsator. So far, only one eclipse has been detected, but future data releases might reveal more eclipses. This object was classified as $\gamma$ Dor star, and the eclipse was detected using our extractor method. The amplitude spectrum clearly shows several significant frequencies in the range 1-2 $d^{-1}$. The total time span of the data is yet too short to have a sufficient frequency resolution for asteroseismological studies.}
\label{KIC3867593}
\end{figure*}

\begin{figure*}
 \centering
   \includegraphics[width=16cm,angle=270,scale=0.7]{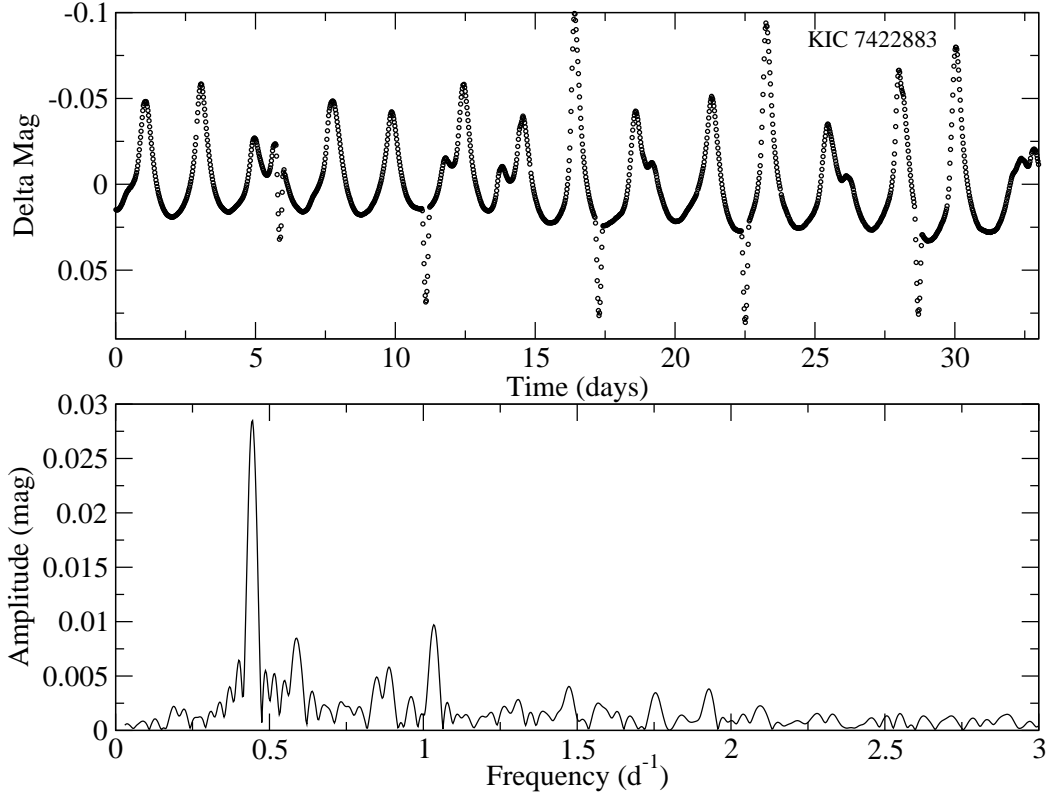}
\caption{An eclipsing binary featuring both primary and secondary eclipses, and additional variability of SPB or $\gamma$ Dor type, with a main frequency of $0.44$ $d^{-1}$. From the distance between the eclipses, we can see that it concerns an eccentric system. Note the similarity with KIC 11285625 in \cite{Gilliland-Kepler}. This object was classified as $\gamma$ Dor by our regular classifier, and the eclipses were detected with our extractor method.}
\label{KIC7422883}
\end{figure*}

\begin{figure*}
 \centering
   \includegraphics[width=16cm,angle=270,scale=0.7]{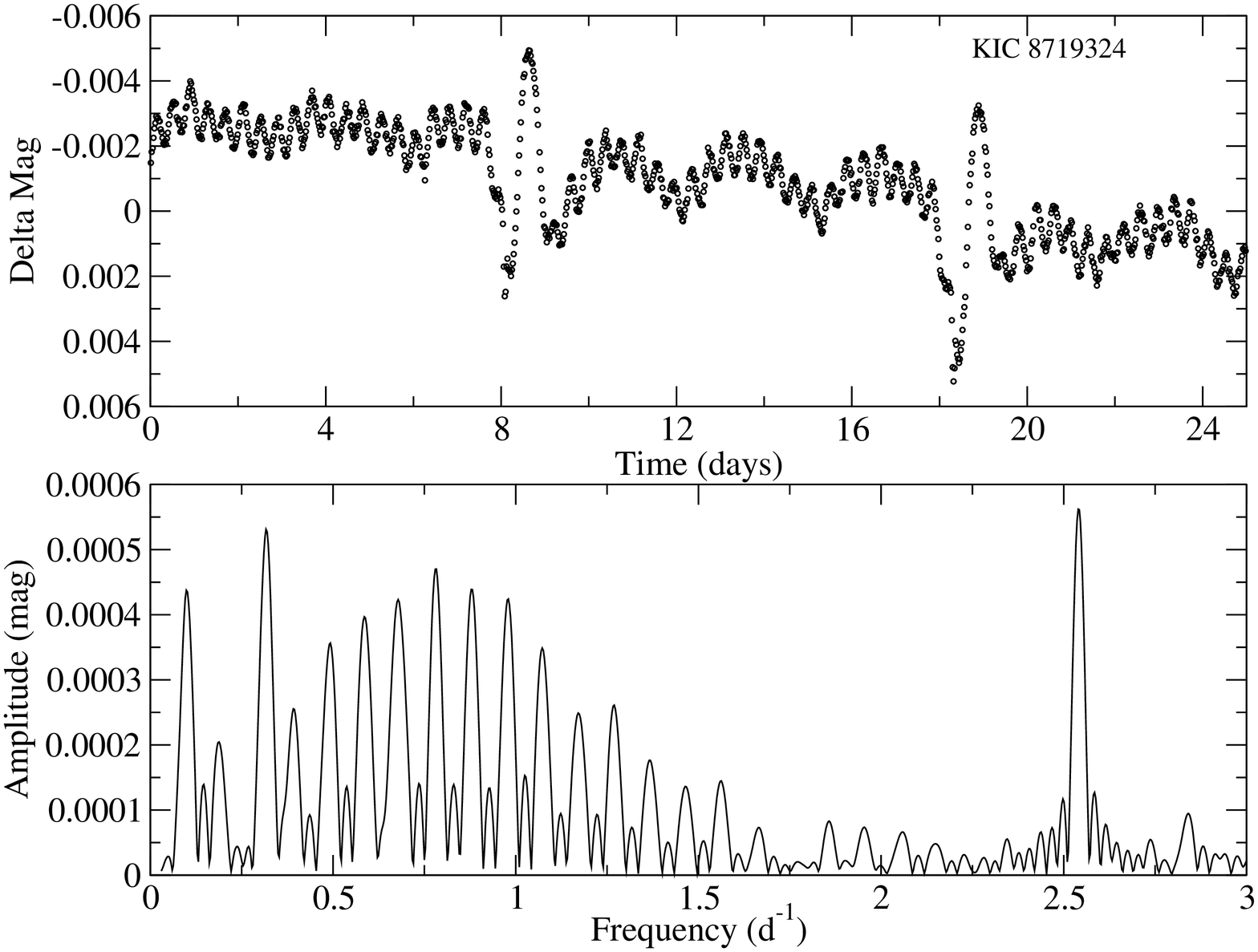}
\caption{An intriguing light curve, showing several phenomena at the same time: eclipses followed by a sudden and short-lived increase in brightness, modulation of the light curve at a period that might be a subharmonic of the orbital period, and additional (pulsational) variability at shorter timescales. The unusual combination of variability at different timescales, caused this object to be assigned to the stellar activity class with low probability (see further for a description of this class), but the eclipses were detected using our extractor method.}
\label{KIC8719324}
\end{figure*}

We have detected several objects showing both clear eclipses and additional variability in their light curves. In some cases, multiperiodic variability is present, indicative of SPB, $\gamma$ Dor or $\delta$ Sct type non-radial pulsations. Some nice examples are shown in Figs. \ref{KIC3867593}-\ref{KIC8719324} and another one from the KASC sample (KIC 11285625) was already shown in \cite{Gilliland-Kepler}. Those objects deserve our special attention, and for some of them, spectroscopic follow-up is planned. Combining the \textit{Kepler} light curves with ground-based spectra, it is possible to derive the orbital elements of the binary system, and model-independent estimates of stellar masses and radii. Those are key parameters needed for asteroseismological studies, and they are difficult to derive otherwise.

The light curves of those systems are difficult to identify in a single step with a supervised classifier, since different phenomena are present at the same time, and their relative strengths can vary a lot. For example, a light curve with eclipses and additional pulsations, will be classified as pulsating variable if the amplitude of the pulsation(s) in the Fourier spectrum is larger than the amplitude of the orbital peaks due to the eclipses (e.g. for KIC7422883). The reverse situation will cause the light curve to be classified as being of the eclipsing binary type. The situation is not that clear-cut, when both phenomena have comparable strength in the Fourier spectrum. The current version of the classifier takes the 3 most significant frequency peaks into account (each with a maximum of 4 harmonics), and it can happen that the first one is related to the pulsations, while the others are related to the eclipses. This confuses the classifier, certainly if the orbital period is very different from the pulsation period(s) (e.g. not in the same range as the typical pulsation periods for the type of variable present in the binary system). In these cases, the class labels have to be treated with caution (e.g. for KIC8719324). Therefore, we used the results of our eclipsing binary extractor to complement the classification results to detect those objects.

In total, we could identify about 14 candidate pulsators in eclipsing binary systems. Of those, 5 are classified as SPB or $\gamma$ Dor and indeed show pulsations of that type. They are flagged by the extractor method, indicating the presence of at least one eclipse in the high-pass filtered version of the light curve. Five objects have been identified as eclipsing binary by the classifier, and the additional variability was discovered by visual inspection of the binary sample. The remaining 4 objects are flagged by the binary extractor method (but not recognized as binary by the normal classifier), and the additional variability was again discovered by visual inspection.

\subsection{Samples of candidate non-radial pulsators}

Since we find a large number of new candidate non-radial pulsators, we have examined some group properties of the samples. Our classifiers only use information obtained from the \textit{Kepler} light curves (white light), and therefore cannot reliably discriminate between $\delta$ Sct and $\beta$ Cep pulsators, nor between SPB and $\gamma$ Dor pulsators. Their pulsation spectra are often very similar, but their positions in the Hertzsprung-Russell diagram are very different. We therefore use the 2MASS magnitudes from the KIC catalogue to analyse the samples in more detail. To better answer the question of how many stars are truly good candidate members for those 4 classes, we have compared the 2MASS colour indices of the stars in our sample with those of bona-fide class members from the literature. In fact, we first determined the observational instability domain of those classes in 2MASS colour space. For the $\beta$ Cep class and the SPB class, we used the extensive tables compiled by P. De Cat (available at \url{http://www.ster.kuleuven.be/~peter/Bstars/}), for the \break $\delta$ Sct class, we used the catalogue by \cite{Rodriguez-DSCUT}, and for the $\gamma$ Dor class, we used the lists presented in \cite{Cuypers-GDOR}, \cite{Aerts-GDOR} and \cite{Handler-GDOR}.

For each of the combinations $\beta$ Cep/$\delta$ Sct and SPB/$\gamma$ Dor, we made 2MASS J-H versus H-K colour plots showing both the bona-fide literature samples and the candidate \textit{Kepler} samples we obtained with our classifiers. We have first cleaned our samples to retain only the best candidates, to see if they fall in the regions occupied by known class members. Cleaning has been done by imposing limits on the Mahalanobis distance to the training class center, as described in Sect. 4.2.

We have also investigated the interstellar reddening in the \textit{Kepler} field, to check whether significant colour shifts are present and might hamper our conclusions. The effects of interstellar absorption are relatively small for the H, J and K infrared photometric bands. We estimated E(H-K) and E(J-H) for the majority of the \textit{Kepler} stars, by using the derived extinction values $A_V$ from the \textit{Kepler} field description available on the NASA MAST archive (Multimission Archive at STScI), in combination with the ratios $A_{band}/A_{V}$ presented in \cite{Rieke}. In Figs. \ref{spb-gdor} to \ref{rot} and Figs. \ref{rot-sr} to \ref{act}, the average reddening vectors for the \textit{Kepler} samples are indicated by means of a black arrow. For every star with available extinction values, we estimated E(H-K) and E(J-H). The components of the reddening vector are then constructed by taking the sample average of E(H-K) and E(J-H). Typically, the standard deviations of E(H-K) and E(J-H) for each \textit{Kepler} sample are only about one third of their average values, thus justifying that we only show the average reddening vectors. We did not estimate the reddening for the bona-fide literature samples, since these samples contain nearby objects (mainly measured by HIPPARCOS) and are less influenced by reddening compared to the \textit{Kepler} samples.

Figure \ref{spb-gdor} shows the results for the SPB/$\gamma$ Dor classes. Clearly, most of our candidates fall nicely within the expected colour region of the $\gamma$ Dor class, taking the effects of reddening into account. This is not surprising, given that $\gamma$ Dor stars are less massive (1.5 to 1.8$M_\odot$) compared to SPB stars (2 to 7 $M_\odot$), hence much more abundant according to the initial mass function \citep[see e.g.][]{Scalo-IMF}. Given that we did not take any colour information into account to classify the stars, this is a very nice result, showing that these classes of non-radial pulsating stars can be identified reliably using well-sampled white-light photometric time series. We conclude that our method can separate SPB/$\gamma$ Dor candidates from other variability types, and that we need colour information only in a second stage, to discriminate between SPB and $\gamma$ Dor. We should also find at least some SPB candidates, given the large sample of stars. Indeed, a few of our candidates fall within the SPB domain in colour space and are likely SPB stars. Their visual magnitudes also do not exclude their SPB nature: these are bright objects and can only be present at the bright end of the \textit{Kepler} sample.

Figure \ref{dscut-bcep} shows a similar plot for the $\beta$ Cep/$\delta$ Sct classes. The majority of our $\delta$ Sct candidates falls within the expected colour region for this variability class, taking the effects of reddening into account. This illustrates again the quality of our classification based on a single \textit{Kepler} light curve only. We don't expect to find many \break $\beta$ Cep candidates, again based on the initial mass function ($\beta$ Cep stars have masses in the range 8-18$M_\odot$, while $\delta$ Sct stars have masses in the range 1.5-2.5$M_\odot$), but also given their high luminosities: most \textit{Kepler} targets are too faint to be $\beta$ Cep stars, they would have to be at a distance placing them outside the Milky Way! We should find even less $\beta$ Cep than SPB candidates. Only one or two of our candidates fall nicely in the $\beta$ Cep domain, and the visual magnitude is within the range of those of known galactic $\beta$ Cep stars. Their light curves indeed show clear pulsations with frequencies in the $\beta$ Cep range, making them convincing candidates. Spectroscopic observations are required to confirm their nature, also for the few SPB candidates we find.

About 4000 objects in the \textit{Kepler} Q1 public dataset are present in the asteroseismology dataset as well \citep[KASC, see][]{Gilliland-Kepler}. We have also checked how many of our candidates are present in the corresponding class lists of the asteroseismology dataset, since candidate lists of these variables were made prior to the \textit{Kepler} mission (the objects in the KASC dataset are distributed over several working groups, according to their suspected variability type). We found that only 36 out of our 295 best SPB/$\gamma$ Dor candidates, and 75 out of our 313 best $\beta$ Cep/$\delta$ Sct candidates, are present in the asteroseismology dataset. None of the 36 $\gamma$ Dor candidates was present in the $\gamma$ Dor sample of the asteroseismology data, while 55 out of the 75 $\beta$ Cep/$\delta$ Sct candidates are present in the $\delta$ Sct sample. Clearly, we find many more good candidate pulsators in the public dataset, which are not present in the asteroseismology set. Note that by imposing less stringent limits on the Mahalanobis distance to the class centres, our sample sizes even increase, but we would include less obvious candidates, and more false-positives. This results is an increased scatter of the sample in 2MASS colour space. The border cases are also of interest, however, since we expect to find them at the borders of the pulsational instability strips, thus helping to better constrain them. Longer Kepler time-series of these stars are thus of immense importance for a better understanding of stellar structure and evolution.

\begin{figure*}
 \centering
   \includegraphics[width=16cm,angle=270,scale=0.6]{spb-gdor-new.ps}
\caption{Comparison in 2MASS colour space of samples of bona-fide SPB and $\gamma$ Dor stars, and the candidates we find in the \textit{Kepler} data. The blue star symbols represent bona-fide SPB stars, red squares represent bona-fide $\gamma$ Dor stars, and the green triangles represent our sample of candidate SPB/$\gamma$ Dor stars. The estimated average reddening vector for the \textit{Kepler} sample is indicated with the black arrow. Most of our candidates fall within the observational $\gamma$ Dor colour region and are likely to be good candidates. Only few objects are good SPB candidates, as expected, since SPB stars are much less abundant.}
\label{spb-gdor}
\end{figure*}

\begin{figure*}
 \centering
   \includegraphics[width=16cm,angle=270,scale=0.6]{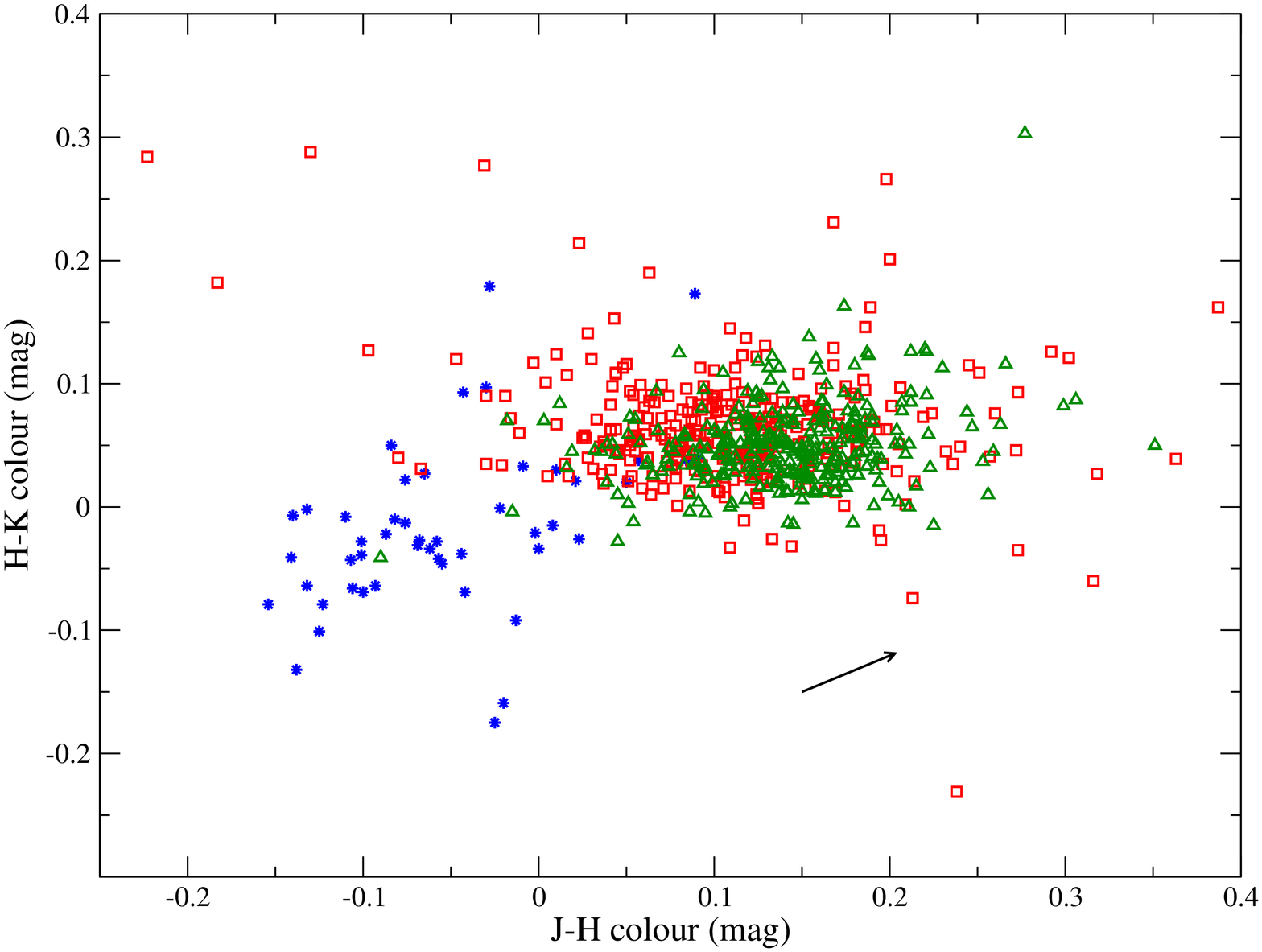}
\caption{Comparison in 2MASS colour space of samples of bona-fide $\delta$ Sct and $\beta$ Cep stars, and the candidates we find in the \textit{Kepler} data. The blue star symbols represent bona-fide $\beta$ Cep stars, red squares represent bona-fide $\delta$ Sct stars, and green triangles represent our \textit{Kepler} sample of $\beta$ Cep/$\delta$ Sct stars. The estimated average reddening vector for the \textit{Kepler} sample is indicated with the black arrow. As can be seen, most of them fall within the colour boundaries of the observational $\delta$ Sct instability domain, and are likely to be good candidates. Only one or two candidates are situated in the $\beta$ Cep region, and are good $\beta$ Cep candidates.}
\label{dscut-bcep}
\end{figure*}

\subsection{Rotational modulation and stellar activity}

Both the rotational modulation and stellar activity classes are recent additions to our training set, and it is therefore important to assess how well these variability types can be distinguished from the many other forms of stellar variability. Cross-validation tests performed on our training set, show that we can distinguish those light curves well from the other training classes. However, to check the real performance of the classifier for these new classes, a large and completely independent data set has to be used. Since the \textit{Kepler} Q1 public data contain more than 150\,000 light curve of excellent quality, and are expected to contain many objects showing the signatures of activity and rotational modulation, this is an ideal dataset for this purpose.

\begin{figure*}
 \centering
   \includegraphics[width=16cm,angle=270,scale=0.6]{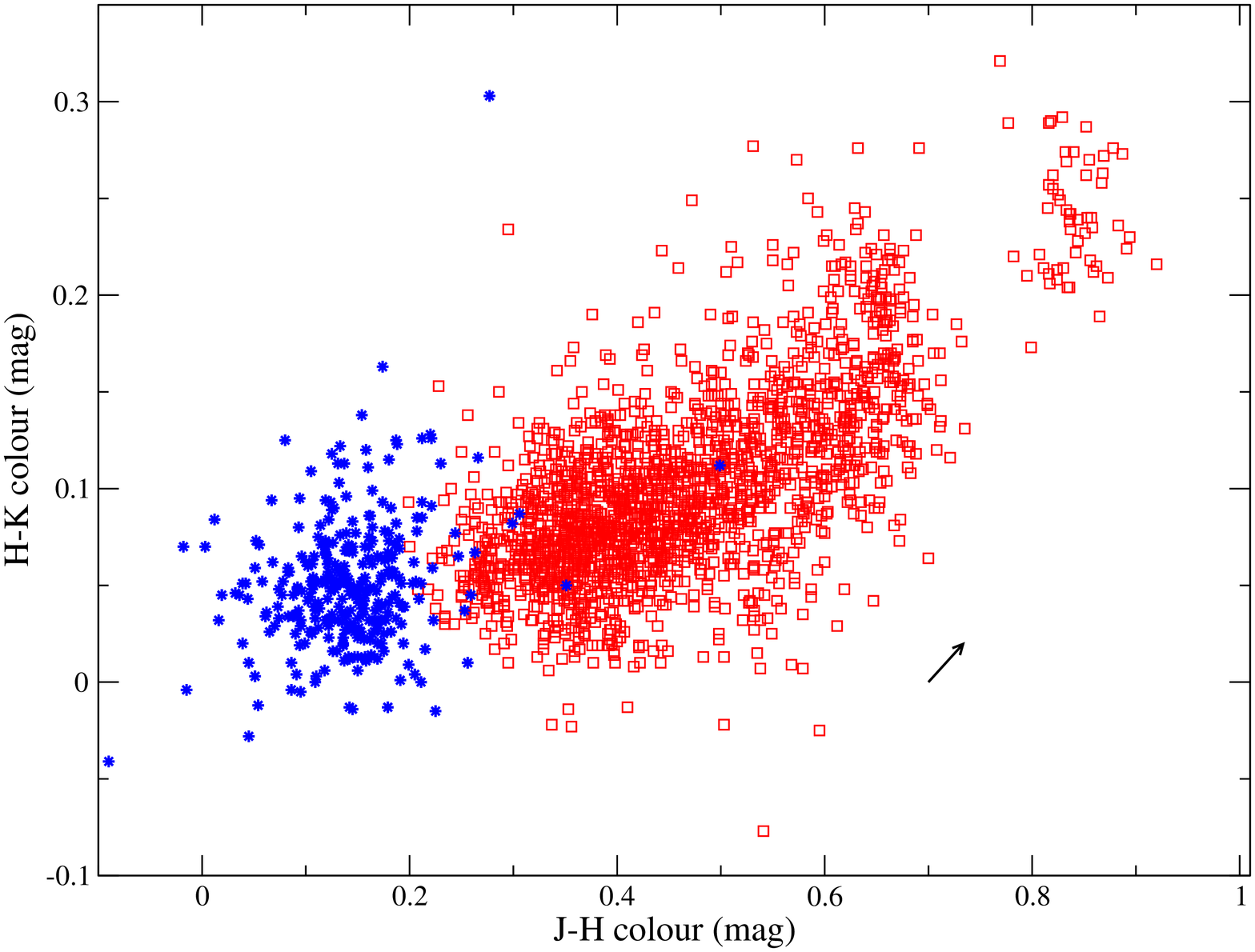}
\caption{Objects assigned to the rotational modulation class, plotted in 2MASS colour space (red squares). For comparison, we also plotted our \textit{Kepler} $\delta$ Sct sample (blue stars), the same as shown in Fig. \ref{dscut-bcep}. The estimated average reddening vector for the \textit{Kepler} rotational modulation sample is indicated with the black arrow. A clear subgroup of the rotational modulation sample can be distinguished at the redest colour part of the plot. We have visually checked several light curves of objects located in both subgroups of the sample, two typical examples are shown in Fig. \ref{rot-example}. }
\label{rot}
\end{figure*}

\begin{figure*}
 \centering
   \includegraphics[width=16cm,angle=270,scale=0.7]{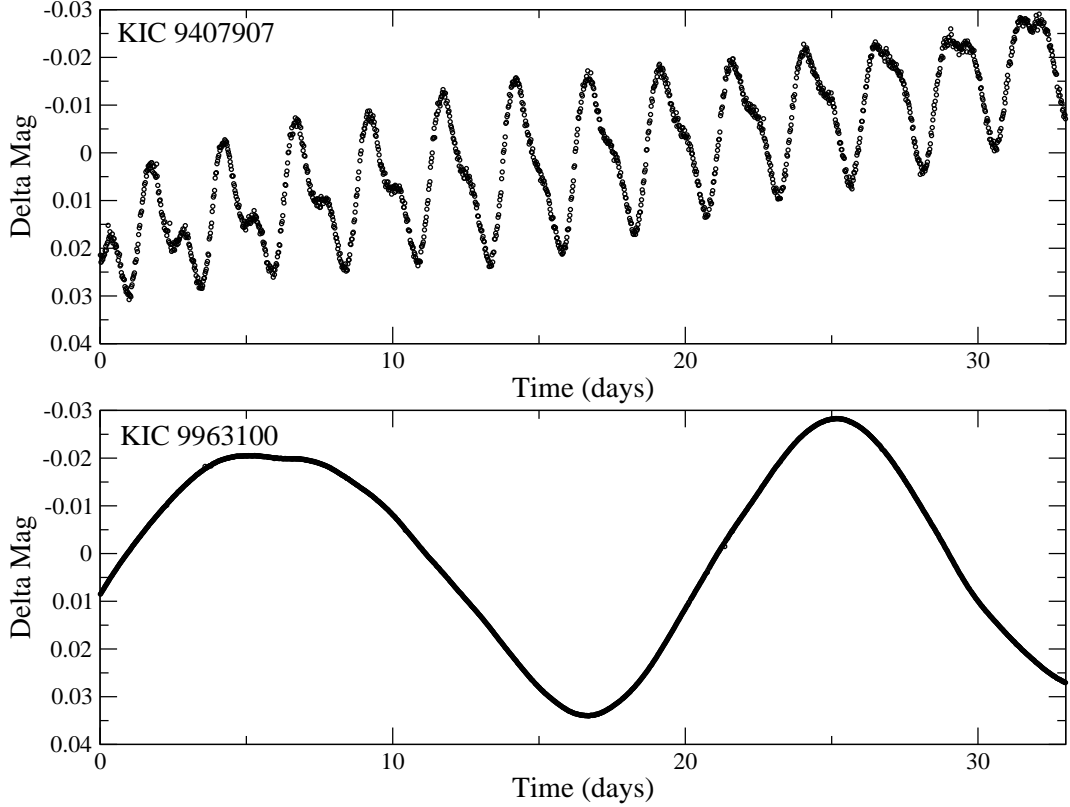}
\caption{Two examples of \textit{Kepler} light curves of objects assigned to the rotational modulation class, but clearly occupying a different region in 2MASS colour space (see Fig. \ref{rot}). The first example belongs to the biggest subgroup and clearly shows the signatures of rotational modulation, as do most of the examples in this subgroup. The second example belongs to the small subgroup. Most examples there show similar lightcurves, with long periods or trends and large-amplitude variability, resembling pulsations of semi-regular variables.}
\label{rot-example}
\end{figure*}

\begin{figure*}
 \centering
   \includegraphics[width=16cm,angle=270,scale=0.6]{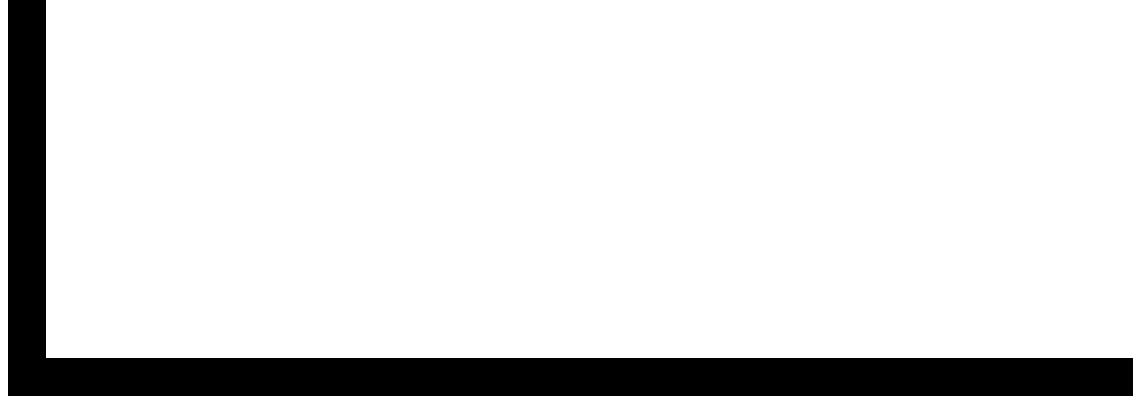}
\caption{The same plot as Fig. \ref{rot}, but now with the HIPPARCOS sample of semi-regular variables shown as well (green triangles). The estimated average reddening vector for the \textit{Kepler} rotational modulation sample is indicated with the black arrow. The small subgroup of our Kepler rotational modulation sample is clearly situated in the colour region occupied by the semi-regular variables.}
\label{rot-sr}
\end{figure*}

We made a selection of the best rotational modulation candidates, again by imposing limits on the Mahalanobis distance to the class center. Using a cutoff-value of 1.5, we still retain almost 2000 candidates. Those candidates are plotted in 2MASS colour space in Fig. \ref{rot}. For comparison, we also plot the same $\delta$ Sct sample as shown in Fig. \ref{dscut-bcep}, to show where the sample is located in the colour diagram with respect to the other classes. We can see that the rotational modulation sample is very well separated from the pulsator classes, while we did not use any colour information in the classification process. Also apparent is the clear subgroup visible in the upper right corner of the diagram. We have visually checked several light curves of objects located in both subgroups, revealing that these two subgroups really contain different kinds of objects. Typical examples of both groups are shown in Fig. \ref{rot-example}, illustrating the periodicity in the light curves. Many objects in the biggest group show very clear signs of rotational modulation in their light curves, similar to the CoRoT light curves in the training set for this class. The objects in the small subgroup all exhibit clear long-term variability, with relatively large amplitudes. These are very red objects, and some of the light curves resemble those of semi-regular variables. In Fig. \ref{rot-sr}, we compare the position of these objects in the colour diagram to the regions occupied by semi-regular variables detected by the HIPPARCOS mission \citep{HIPPARCOS}. The small rotational modulation subgroup clearly falls within the semi-regular region. They are not classified as semi-regular variables with our methods, but this is due to the insufficient time-span of the current light curves. Further investigation and time series of a longer time span are needed to shed more light on this group of variables.

The inclusion of this new class clearly constitutes an improvement of our classification capabilities, since many of those variables are present, and they can now be recognized well. Most of our candidates are located in the cool regions of the colour diagram, where we expect to find those stars. We do see some contamination of red giant stars in the sample of candidates though, suggesting that we need to tweak the classification parameters to avoid this confusion. With the many good example light curves present in the \textit{Kepler} data, we plan to improve the definition of this class.

\begin{figure*}
 \centering
   \includegraphics[width=16cm,angle=270,scale=0.7]{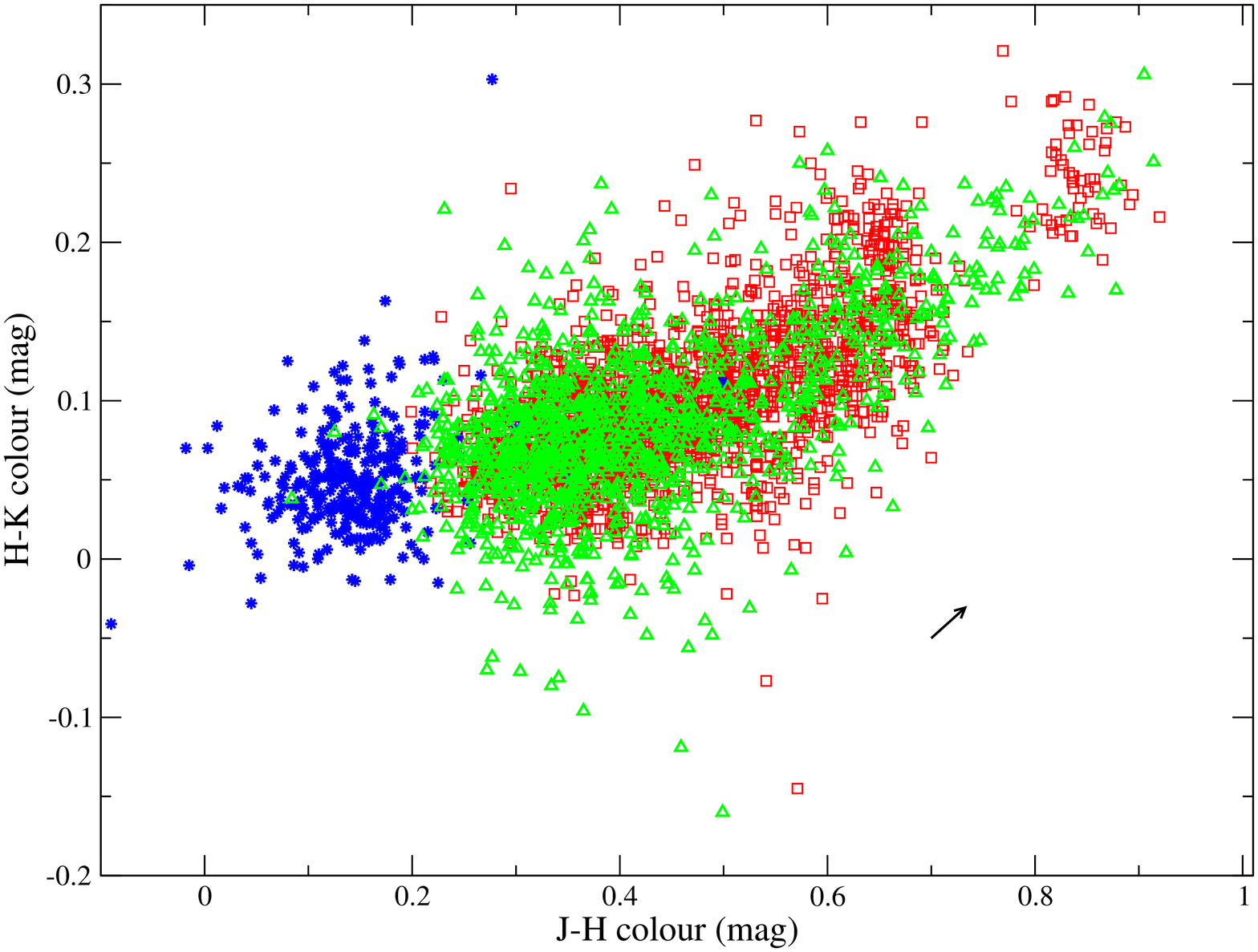}
\caption{The same plot as Fig. \ref{rot}, now with our \textit{Kepler} sample of objects assigned to the stellar activity class overplotted (green triangles). The estimated average reddening vector for the \textit{Kepler} stellar activity sample is indicated with the black arrow. Both the stellar activity and the rotational modulation samples mainly occupy the same region in colour space, corresponding to cool main-sequence stars.}
\label{act}
\end{figure*}

\begin{figure*}
 \centering
   \includegraphics[width=16cm,angle=270,scale=0.7]{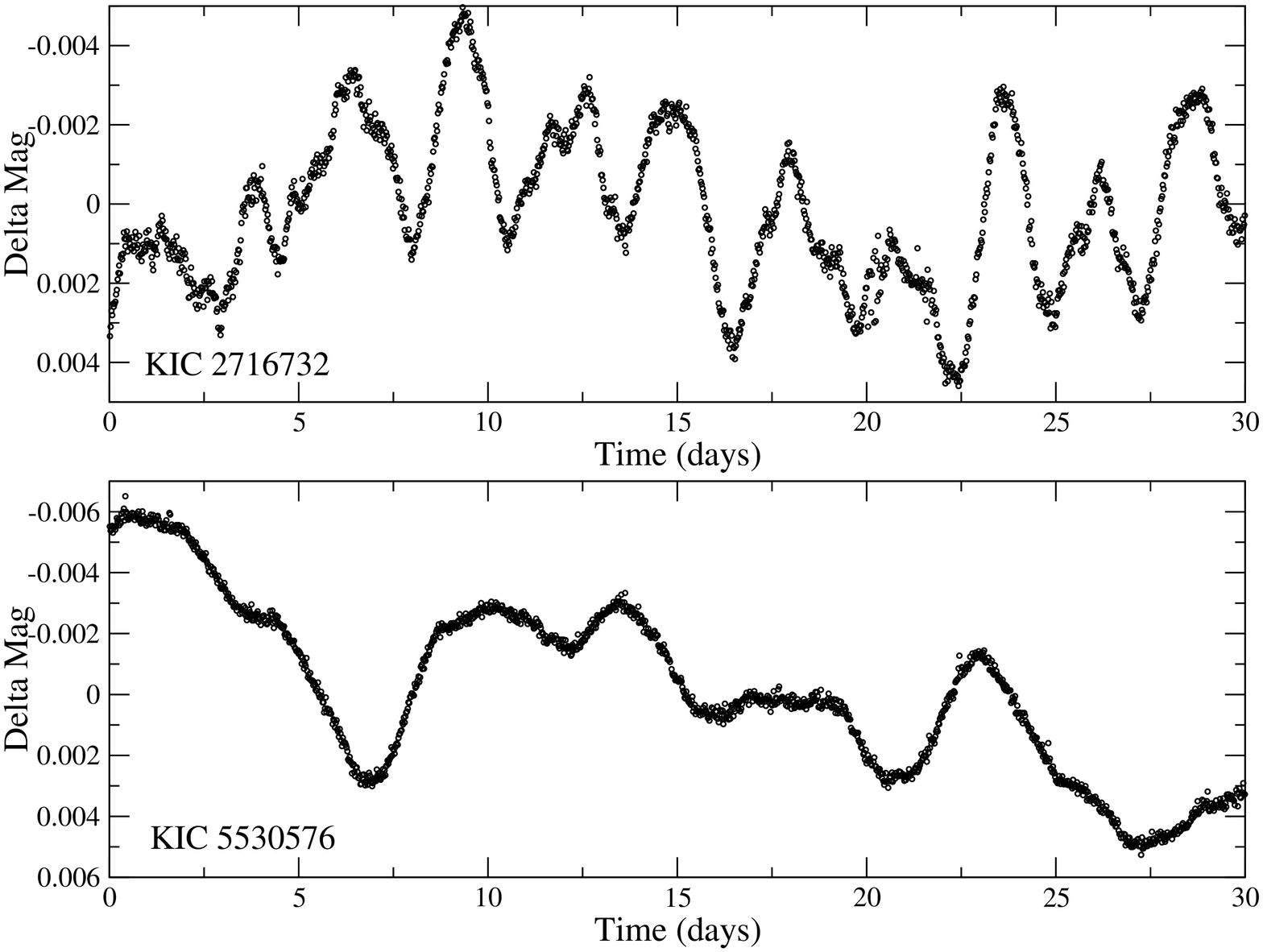}
\caption{Two example of \textit{Kepler} light curves of objects assigned to the stellar activity class. They exhibit variability which is not strictly periodic and with relatively long quasi-periods and low amplitudes.}
\label{act-examples}
\end{figure*}

For the stellar activity class, we used a similar limit on the Mahalanobis distance to the class centre to select the best \textit{Kepler} candidates, retaining about 1200 objects. Note that more than 19\,000 objects are assigned to this class in total, not surprising given the expected abundances of active main-sequence stars in the \textit{Kepler} sample. Figure \ref{act} shows the position of the best candidates in 2MASS colour space. The $\delta$ Sct and rotational modulation candidates are shown as well, for comparison. The activity sample occupies the same region in colour space as the rotational modulation sample, corresponding to cool main-sequence stars. We indeed expect to find many active stars in this region. Note also that the activity sample is very well separated from the pulsator classes in colour space, again without using colour information in the classification process. Figure \ref{act-examples} shows two typical examples of light curves that ended up in the activity class. They show rather irregular variability (compared to the rotational modulation class) with long periods and small amplitudes, similar to the CoRoT light curves in the training set. Some objects in the sample show stricter periodic light curves, similar to those assigned to the rotational modulation class. The differences between those classes are based on light-curve morphology rather than on astrophysical grounds, since stellar activity and rotational modulation due to spots are related phenomena, occurring for stars in the same regions of the Hertzsprung-Russell diagram. Therefore, overlap between those classes is present, but the more regular light curves, with clear signs of stellar spots, will end up in the rotational modulation class. We believe it is useful, however, to keep the subdivision, since the light curves of both subclasses can have a very different morphology. Mixing those together in one class would degrade the classification performance. Another reason to keep the subdivision is the fact that we can reliably identify the `simpler' light curves showing clear signs of rotational modulation, the latter being better suited to study, e.g., stellar rotation and perform spot modeling.


\section{Comparison with TrES ground-based data}
Part of \textit{Kepler's} field-of-view overlaps with one of the fields observed by the ground-based TrES survey (Trans-Atlantic Exoplanet Survey). The goal of this survey was the detection of transiting planets, using a network of three ten-centimetre optical telescopes. About 26\,000 TrES light curves of the overlapping \textit{Lyr1} field have been analysed using adapted automated classification methods (Blomme et al. submitted to MNRAS). We have done a cross-matching based on coordinates and magnitudes to identify common objects in both the TrES and \textit{Kepler} datasets. Using a maximum search radius of 2 arcmin, we found 9963 matches.
It is interesting to compare the quality of the light curves and to see how well the classifiers performed on data having a much higher noise-level and containing daily gaps due to the day-night rhythm, in view of future ground-based surveys containing time series data. 
For the 9963 matching objects, we compared the dominant frequency detected in the TrES light curve with the one detected in the \textit{Kepler} light curve. Frequencies are taken to be equal if their difference is smaller than the frequency resolution obtainable with the \textit{Kepler} light curves: $\mid f_{1,Kepler}-f_{1,TrES}\mid$ $<$ 1/T, with T the total time span of the \textit{Kepler} light curves (note that the time span of the TrES light curves is about twice that of Kepler). We only considered frequencies higher than 0.6 $d^{-1}$, to assure that $f/\Delta f$ is sufficiently large (minimal value of $\sim$20). This way, we found 119 confident frequency matches amongst the 9963 common objects in both databases. We then checked how many of those got assigned the same variability class by our classifiers. The results are summarized in Table \ref{TRES-Kepler}. In total, 48 out of the 119 objects are assigned to the same class. The remaining 71 objects are assigned either to different classes, or classified as `MISC' (Miscellaneous) in both cases. Amongst those are also 4 objects classified as $\delta$ Sct from the TrES data, and classified as eclipsing binary from the \textit{Kepler} data. It concerns short period binaries whose orbital period is in the same range as the pulsation periods of typical $\delta$ Sct stars. These illustrate that the higher quality of the Kepler data improved the classification of those targets (they turn out to be binaries indeed).

\begin{table}
\caption{Number of variables whose classification from TrES and \textit{Kepler} are equal and the variability class they are assigned to.}
 \begin{tabular}{l|l}
  \hline
  Stellar variability class & \# identical classifications \\
  \hline
$\delta$ Sct stars&34\\
Binaries (eclipsing and ellipsoidal)&8\\
$\gamma$ Dor stars&5\\
RR Lyr stars, subtype c&1\\
 \hline
\end{tabular}
\label{TRES-Kepler}
\end{table}

\begin{figure*}
 \centering
   \includegraphics[width=16cm,angle=270,scale=0.8]{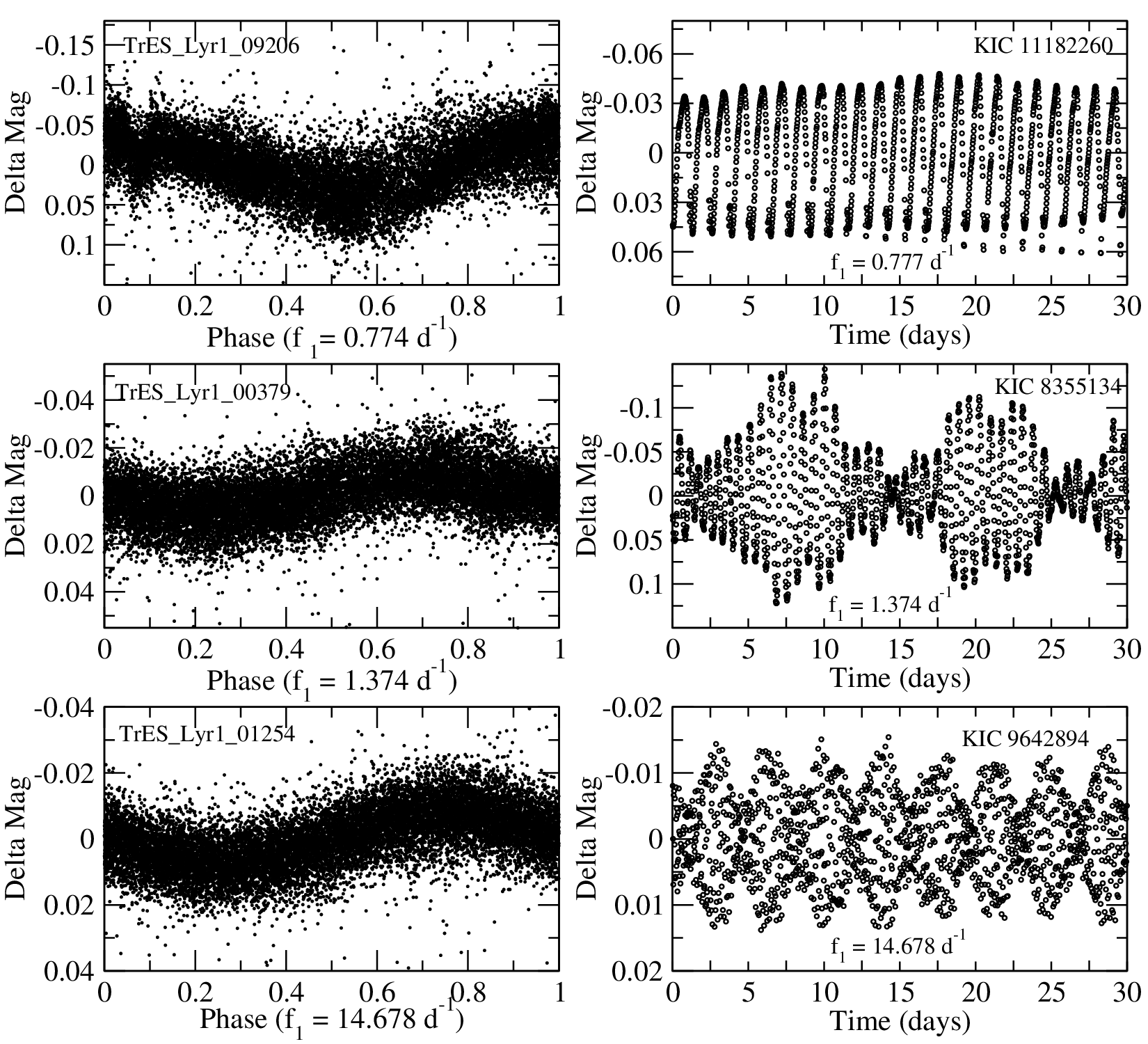}
\caption{Some examples of objects for which we found a match between the \textit{Kepler} and TrES data. From top to bottom, the TrES and Kepler light curves of, respectively: an eclipsing binary, a $\gamma$ Dor candidate, and a $\delta$ Sct candidate. The TrES light curves have been phased according to the dominant frequency found in the data, for visibility reasons.}
\label{tres}
\end{figure*}

Figure \ref{tres} shows some examples of variables whose classification from TrES and \textit{Kepler} are equal: both the TrES and the \textit{Kepler} light curves are plotted for an eclipsing binary with additional variability, a candidate $\gamma$ Dor pulsator and a candidate $\delta$ Sct pulsator. The TrES light curves have been phased according to the dominant frequency found in the data, for visibility reasons. These ground-based data have a much poorer quality than the \textit{Kepler} data, and the variability is very difficult to see by eye in the original light curve. Nevertheless, those objects were classified correctly using the TrES data, showing the robustness of our methods. For the pulsating stars, we also find exactly the same dominant frequency peaks in the TrES and \textit{Kepler} light curves, showing the reliability of the frequencies and the stability of the pulsation modes. The latter is very useful when doing asteroseismological studies of individual objects, since analysing two independent datasets is the best way to have certainty about detected frequencies. Obviously, the \textit{Kepler} data allow many more significant frequencies to be detected than the TrES data do, but we could at least verify the reliability of the three most dominant frequencies in ground-based data which were assembled with a completely different goal than asteroseismology, and, along with it, the suitability of target selection for follow-up dedicated studies of the best class candidates.

\section{Conclusions}

We have presented a global variability study of the public \textit{Kepler} data, measured in the first quarter of the mission. In total, we analysed more than 150\,000 light curves using automated classification and extractor methods. This database is unprecedented, never before did we have access to such a large sample of very-well sampled light curves with such a high photometric precision. It is therefore an excellent dataset to perform statistical studies on the relative class populations of variable stars of all kinds and better constrain their instability domains.

To improve our detection capabilities, we have introduced two `new' classes in our classification scheme: variables showing rotational modulation and active stars. We have also supplemented our classifiers with a dedicated extracted method for eclipsing binaries, to improve the detection of faint eclipses and eclipses in light curves containing other variability as well. The method proves to be very effective, we could significantly increase the number of detected binaries, and identified several pulsating stars in eclipsing binary systems. The latter are of special interest in the field of asteroseismology.

We presented variability estimates and class statistics, which are compared with similar studies done on the CoRoT exoplanet data. The results for the relative class populations are rather similar, since both missions focus on main-sequence targets, with the goal to detect earth-like planets. This implies that the number of classical pulsators such as RR Lyr and Cepheids in the datasets are small, compared to the number of non-radial pulsators such as $\delta$ Sct and $\gamma$ Dor.

The samples of candidate non-radial pulsators we identified have been evaluated by using 2MASS colour indices. We compared the position of our candidates in the colour diagram with those of known class members from the literature. The results convincingly show that our samples contain many new class members for the $\delta$ Sct and $\gamma$ Dor classes, a few very good SPB candidates, and one or two candidates for the $\beta$ Cep class. The use of colour indices allowed us to discriminate between $\delta$ Sct /$\gamma$ Dor and SPB/$\beta$ Cep respectively, while this is not possible using only the light curve information.
Our classifiers are, however, very well capable of separating those combinations of two classes from other variability types, as confirmed by the well constrained regions the candidates occupy in colour space.

We have positively evaluated the performance of our classifiers for the new rotational modulation and activity classes. Many good candidates could be identified for both classes, and they occupy well-defined regions in colour space, corresponding to cool main-sequence stars. This is where we expect to find these types of variability. We also discovered a clear subgroup in our rotational modulation sample, containing redder objects whose light curves show long-term variability. The nature of these objects needs to be investigated further, but we have strong indications that they are semi-regular variables.

Future work includes more detailed object studies and spectroscopic follow up of selected non-radial pulsators, and especially pulsating stars in eclipsing binary systems. We also plan to keep updating our training set used for the supervised classification, by including high quality \textit{Kepler} light curves once the true class membership is confirmed spectroscopically. A detailed clustering analysis of the \textit{Kepler} database, as in \cite{Sarro-clustering} for the CoRoT data, is planned as well, since it contains such a large number of excellent light curves. When more \textit{Kepler} data will be released, we will have access to longer time-series for most objects, allowing us to identify and study long-period variables as well.

\begin{acknowledgements}
We would like to express our special thanks to the numerous people who helped making the \textit{Kepler} mission possible.
The research leading to these results has received funding from the European
Research Council under the European Community's Seventh Framework Programme
(FP7/2007--2013)/ERC grant agreement n$^\circ$227224 (PROSPERITY), from the
Research Council of K.U.Leuven (GOA/2008/04), from the Fund for Scientific
Research of Flanders (G.0332.06), and from the Belgian federal science
policy
office (C90309: CoRoT Data Exploitation, C90291 Gaia-DPAC).
This publication makes use of data products from the Two Micron All Sky Survey, which is a joint project of the University of Massachusetts and the Infrared Processing and Analysis Center/California Institute of Technology, funded by the National Aeronautics and Space Administration and the National Science Foundation.
Some/all of the data presented in this paper were obtained from the Multimission Archive at the Space Telescope Science Institute (MAST). STScI is operated by the Association of Universities for Research in Astronomy, Inc., under NASA contract NAS5-26555. Support for MAST for non-HST data is provided by the NASA Office of Space Science via grant NNX09AF08G and by other grants and contracts.
This research has made use of the SIMBAD database,
operated at CDS, Strasbourg, France.
\end{acknowledgements}

\bibliographystyle{aa}
\bibliography{Kepler-classif-JD.bib}

\appendix
\section{Stellar variability classes}
\begin{table}
\caption{The different variability classes considered by the current version of our supervised classification method. The `Miscellaneous' category stands for objects not belonging to any of the variability classes we consider.}
 \begin{tabular}{l|l}
  \hline
  Stellar variability class & Abbreviation \\
  \hline
$\beta$-Cephei stars&BCEP\\
Classical Cepheids&CLCEP\\
Double-mode Cepheids&DMCEP\\
$\delta$-Scuti stars&DSCUT\\
Eclipsing binaries (all types)&ECL\\
Ellipsoidal variables&ELL\\
$\gamma$-Doradus stars&GDOR\\
Mira variables&MIRA\\
RR-Lyrae stars, subtype ab&RRAB\\
RR-Lyrae stars, subtype c&RRC\\
Double-mode RR-Lyrae stars&RRD\\
RV-Tauri stars&RVTAU\\
Slowly pulsating B-stars&SPB\\
Semi-regular variables&SR\\
Rotational modulation&ROT\\
Active stars&ACT\\
Miscellaneous&MISC\\
 \hline
\end{tabular}
\label{varclasses}
\end{table}

\end{document}